\documentclass[10pt,conference]{IEEEtran}
\IEEEoverridecommandlockouts

\usepackage{graphicx}
\usepackage{stfloats}
\usepackage{textcomp}
\usepackage{xcolor}
\usepackage{xspace}
\usepackage[nointegrals]{wasysym}
\usepackage{pifont}
\usepackage{multirow}
\usepackage{flushend}
\usepackage{caption}
\usepackage{booktabs}
\usepackage{hyperref}
\usepackage{algorithm}
\usepackage{pdfpages}
\usepackage{amssymb}
\usepackage{tabularx}
\usepackage{multicol} 
\hypersetup{
colorlinks = true, 
linkcolor=black, 
citecolor=black, 
urlcolor=black 
}
\usepackage{url}
\usepackage{enumitem}
\usepackage{braket}
\usepackage{amsmath}
\usepackage[breakable]{tcolorbox}
\usepackage{colortbl}
\usepackage{pifont}
\usepackage{float}                  
\usepackage{subfig}                 
\usepackage{overpic}                
\usepackage{algpseudocode}

\usepackage{tikz}
\usetikzlibrary{quantikz}
\usetikzlibrary{positioning}
\usetikzlibrary{arrows.meta}
\usepackage{color}
\usepackage{listings}

\usepackage{times}
\usepackage{graphicx}
\usepackage{epsf}
\usepackage{verbatim}
\usepackage{url}
\usepackage{color}
\usepackage{alltt}

\newcommand{\SmallSpace}{\vspace*{-1.4ex}}

\begin{document}
%
\title{QuanUML: Towards A Modeling Language for Model-Driven Quantum Software Development}

\author{\IEEEauthorblockN{Xiaoyu Guo}
\IEEEauthorblockA{
\textit{Kyushu University}\\
guo.xiaoyu.961@s.kyushu-u.ac.jp}
\and
\IEEEauthorblockN{Shinobu Saito}
\IEEEauthorblockA{
\textit{NTT Computer and Data Science Laboratories}\\
shinobu.saito@ntt.com}
\and
\IEEEauthorblockN{Jianjun Zhao}
\IEEEauthorblockA{
\textit{Kyushu University}\\
zhao@ait.kyushu-u.ac.jp}
}

\maketitle
\begin{abstract}
This paper introduces QuanUML, an extension of the Unified Modeling Language (UML) tailored for quantum software systems. QuanUML integrates quantum-specific constructs, such as qubits and quantum gates, into the UML framework, enabling the modeling of both quantum and hybrid quantum-classical systems. We apply QuanUML to Efficient Long-Range Entanglement using Dynamic Circuits and Shor's Algorithm, demonstrating its utility in designing and visualizing quantum algorithms. Our approach supports model-driven development of quantum software and offers a structured framework for quantum software design. We also highlight its advantages over existing methods and discuss future improvements.
\end{abstract}

\begin{IEEEkeywords}
Modeling Language, Model-Driven Development, Quantum Software, UML
\end{IEEEkeywords}





\section{Introduction}

Quantum computing has emerged as a transformative paradigm, offering computational advantages over classical computing for certain problem classes~\cite{lanyon2010towards,barends2014superconducting,cross2015quantum,benedetti2016estimation,o2016scalable,olson2017quantum}. However, developing reliable quantum software remains challenging due to the intrinsic characteristics of quantum systems~\cite{zhao2020quantum}, including superposition, entanglement, the probabilistic nature of quantum measurements, and the no-cloning theorem~\cite{nielsen2002quantum}. 

With the advancement of quantum algorithms, the use of dynamic circuits is increasing, yet corresponding modeling methods remain limited. Consequently, quantum software development requires new approaches to accurately model, design, and verify quantum systems. These challenges highlight the need for higher-level abstractions to manage the growing complexity of quantum programs.

Model-driven development (MDD) is a proven software engineering methodology in classical computing, where high-level models drive the generation of system implementation \cite{mellor2003model}. MDD helps reduce the complexity of software design by providing an abstraction layer that captures the essential behavior of a system, allowing developers to focus on the core logic rather than low-level implementation details. Applying MDD to quantum software development holds promise for mitigating some of the challenges mentioned above by providing a systematic way to describe, analyze, and generate quantum programs~\cite{gemeinhardt2021towards}. However, current modeling languages, such as the Unified Modeling Language (UML)~\cite{boochunified}, are insufficient to capture the unique properties of quantum computing, thus highlighting the need for an extended modeling language specific to quantum systems.

UML has been widely adopted for classical software development, providing well-defined structures such as class diagrams, sequence diagrams, and state machines~\cite{boochunified}. However, these classical UML constructs are not designed to represent quantum phenomena such as qubits, gates, and entanglement. To address this gap, we propose an extension to UML, which we call \textit{QuanUML}. QuanUML introduces new modeling constructs that allow developers to represent quantum states, operations, and the evolution of quantum systems. By extending UML, we aim to leverage its existing strengths while providing the necessary abstractions for quantum software modeling, ultimately enabling a model-driven approach to quantum software development.

In this paper, we introduce QuanUML, a UML-based modeling language designed specifically for quantum software systems. QuanUML extends the classical UML by introducing constructs that model quantum-specific elements such as qubits, quantum gates, and quantum circuits \cite{nielsen2002quantum}. The language enables developers to model the architecture of quantum systems at a higher abstraction level and quantum algorithms at a specific abstraction level, facilitating the design, analysis, and verification of quantum software systems. Our proposed language provides a bridge between the high-level conceptual modeling of quantum algorithms and the low-level implementations using quantum programming frameworks.

We demonstrate the applicability of QuanUML through detailed case studies of well-known quantum algorithms, such as Efficient Long-Range Entanglement using Dynamic Circuit (\textit{Dynamic Circuit} for short)~\cite{baumer2024efficient} and Shor's Algorithm~\cite{shor1994algorithms}. These case studies showcase how QuanUML can be used to model the structure and behavior of quantum systems, highlighting the benefits of using model-driven development in the quantum domain.

The case studies presented in this paper validate the practical applicability of QuanUML and highlight the modeling challenges associated with different types of quantum algorithms. Efficient Long-Range Entanglement using Dynamic Circuits requires integrating classical control flow into quantum circuits, while Shor's Algorithm necessitates combining high-level and low-level representations to address the complexity of low-level structures. These case studies demonstrate how QuanUML captures key quantum concepts such as entanglement, measurement, and classical control, providing a clear and structured approach to modeling complex quantum operations.

This paper makes the following contributions:
\begin{itemize}[leftmargin=2em] 
\setlength{\itemsep}{2pt}

\item \textit{Extension of UML for Quantum Systems:} We propose QuanUML, an extension of the Unified Modeling Language (UML), specifically designed to model quantum software systems. QuanUML introduces new constructs that capture quantum-specific elements, such as qubits, quantum gates, and quantum operations, enabling more accurate modeling of quantum algorithms and systems.

\item \textit{Model-Driven Quantum Software Development:} QuanUML supports a model-driven approach for quantum software development, allowing for high-level and low-level modeling and automatic code generation. This contribution facilitates the design, analysis, and verification of quantum programs while reducing the complexity of quantum software development.

\item \textit{Case Studies on Quantum Algorithms:} 
We demonstrate the effectiveness of QuanUML through detailed case studies on Dynamic Circuit and Shor's Algorithm. These case studies illustrate how QuanUML models quantum-specific behaviors, such as superposition, entanglement, and measurement, and how it supports the modeling of complex quantum operations.

\item \textit{Improved Abstraction and Scalability:} QuanUML provides a structured framework for modeling quantum software systems at a high level of abstraction. It enables scalability by supporting various quantum algorithms and architectures, making it suitable for diverse quantum computing applications.

\end{itemize}

The rest of the paper is organized as follows. Section~\ref{sec:background} provides background on quantum programming and classical UML. Section~\ref{sec:quanuml} introduces QuanUML, detailing its design and the key extensions made to UML for quantum software systems. Section~\ref{sec:casestudies} presents case studies using QuanUML to Dynamic Circuit and Shor's Algorithm. Section~\ref{sec:relatedwork} discusses related work. Concluding remarks and future research directions are given in Section~\ref{sec:conclusion}.

\section{Background}\label{sec:background}

This section provides the necessary foundation for understanding the context of our work. 

\subsection{Quantum Programming}

\subsubsection{Basic Concepts in Quantum Computing}
Quantum computing represents a fundamental shift from classical computing, utilizing principles such as superposition, entanglement, and quantum interference to solve problems intractable for classical computers. The basic unit of quantum computation is the qubit, which, unlike a classical bit, can exist in a superposition of both \textit{0} and \textit{1} states simultaneously~\cite{nielsen2002quantum}. Another key concept is entanglement, where qubits become interdependent, meaning the state of one qubit can directly influence the state of another, even over large distances \cite{nielsen2002quantum}. Quantum gates operate on qubits to perform computations, and these gates, combined with quantum circuits, form the building blocks of quantum algorithms \cite{nielsen2002quantum}. Quantum measurement, however, collapses the qubit's superposition into a definite state, introducing probabilistic outcomes that are a fundamental aspect of quantum computation.

\subsubsection{Current Quantum Programming Frameworks}
Several quantum programming frameworks have been developed to facilitate the implementation of quantum algorithms on various quantum hardware. Notable frameworks include IBM's Qiskit \cite{ibm2024qiskit}, Google's Cirq \cite{cirq2018google}, and Rigetti's PyQuil \cite{koch2019introduction}. These platforms provide developers with the tools necessary to design and execute quantum circuits, simulate quantum algorithms, and run programs on quantum processors or simulators. 
However, while these frameworks assist in programming and executing quantum algorithms, they lack the high-level abstraction necessary for modeling the structure and behavior of quantum systems in a systematic and visual way.

\subsubsection{Challenges in Modeling Quantum Systems}
Quantum systems pose unique challenges for modeling, mainly due to the probabilistic and non-deterministic nature of quantum mechanics. Classical modeling tools like UML assume deterministic transitions and behaviors, which do not align with the quantum domain where measurement results are probabilistic, and quantum state evolution can be complex and challenging to visualize~\cite{ashktorab2019thinking}. Moreover, quantum algorithms involve phenomena like superposition and entanglement, which further complicate the modeling process. Traditional software engineering tools are insufficient for capturing these quantum-specific behaviors, necessitating the development of new modeling languages and methodologies that can support the unique requirements of quantum computing~\cite{zhao2020quantum}.

\subsection{Classical UML}

\subsubsection{Overview of UML Diagrams}
The Unified Modeling Language (UML) is a standard in classical software engineering for visualizing, specifying, constructing, and documenting the artifacts of software systems \cite{uml2009version}. UML includes a variety of diagram types that represent different aspects of system design. Common diagrams include class diagrams, which depict the static structure of a system; sequence diagrams, which show object interactions over time; and state machine diagrams, which model the dynamic behavior of objects as they transition between states in response to events. These diagrams provide software engineers with a way to communicate system architecture and behavior in a structured and standardized manner.

\subsubsection{Common Applications of UML}
UML has been widely adopted in classical software development for various purposes, including system design, documentation, and communication among stakeholders. For instance, class diagrams are often used to define the relationships between different components in object-oriented programming, while sequence diagrams are employed to model the flow of interactions in distributed systems~\cite{boochunified}. State machine diagrams are frequently used in real-time systems and embedded software to model the different states an object can occupy and the transitions between them~\cite{rurnbaughunified}. The flexibility and expressiveness of UML have made it an indispensable tool in many classical software engineering projects.

\subsubsection{Limitations of UML in Quantum Context}
Despite its widespread use in classical software engineering, UML is not inherently suited to model quantum systems. The main limitation lies in its inability to capture quantum-specific phenomena, such as superposition, entanglement, and the probabilistic nature of quantum measurements~\cite{Perez-Delgado2020quantum,ali2020modeling,perez-castillo2020quantum}. For instance, while state machine diagrams can model transitions between states in classical systems, they do not provide mechanisms for representing the non-deterministic state collapses that occur in quantum systems. Similarly, sequence diagrams do not account for quantum gates or the unique operations that manipulate quantum states. These limitations highlight the need for an extension to UML that incorporates quantum-specific constructs, allowing it to model quantum algorithms and systems effectively.

\section{Quantum Software Modeling using QuanUML}\label{sec:quanuml}
This section introduces our proposed approach, QuanUML, which extends the classical UML to model quantum software systems effectively. 

\begin{figure}
    \centering
    \hspace{-0.7cm}\includegraphics[width=1.06\linewidth]{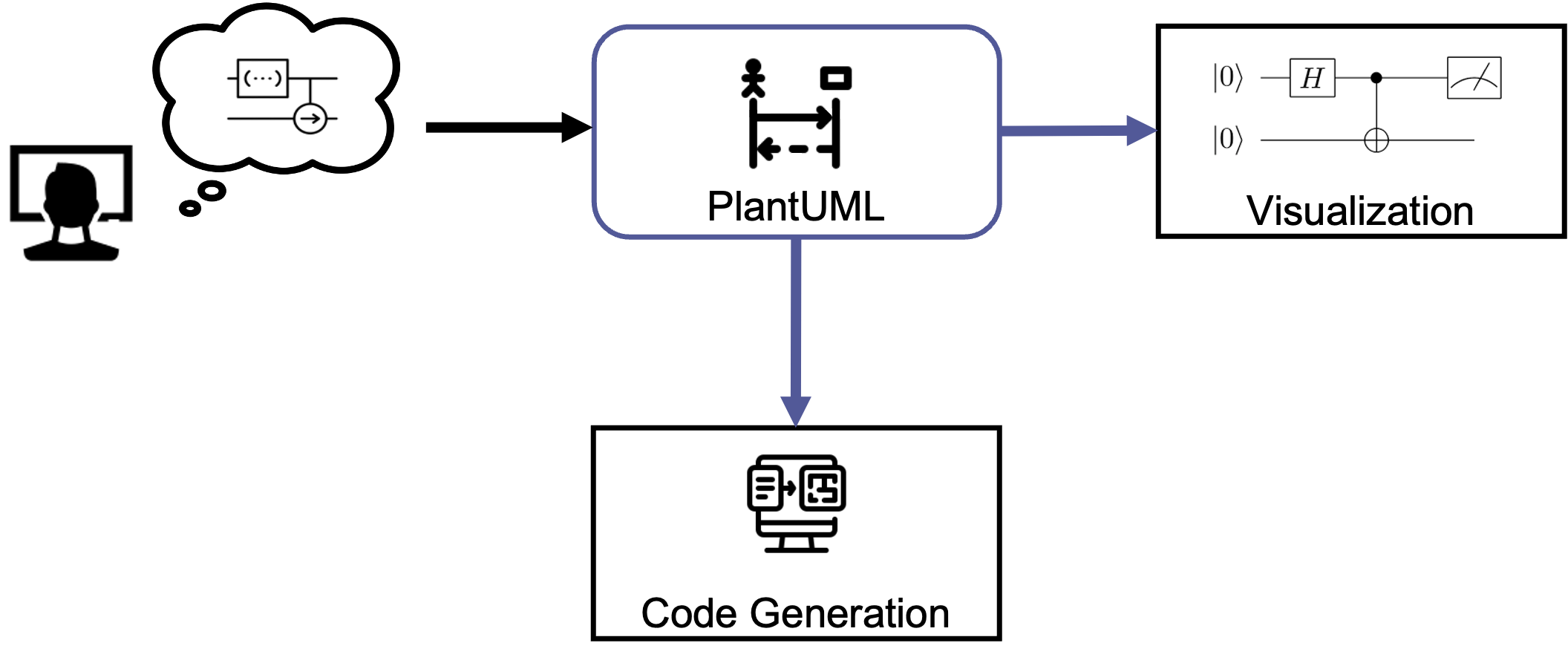}
    \caption{An overview of QuanUML Workflow}
    \label{fig:Workflow}
\end{figure}

\subsection{Overview of QuanUML}
QuanUML is an extension of the Unified Modeling Language (UML) tailored specifically for quantum software systems. While UML has been extensively used in classical software engineering, it lacks the constructs to model quantum-specific elements, such as qubits, quantum gates, and quantum circuits. QuanUML addresses these limitations by introducing new abstractions that capture the behaviors and operations unique to quantum computing. These extensions enable developers to model quantum systems at a high level of abstraction and represent quantum algorithms in detail while maintaining compatibility with UML's classical constructs. This unified approach facilitates the modeling of both quantum and classical components within hybrid systems.

Figure \ref{fig:Workflow} provides an overview of the QuanUML workflow, illustrating the process of modeling quantum systems and generating code. Developers first model the quantum system based on the QuanUML specification. The modeling process consists of two steps: high-level modeling, which captures the quantum system at an abstract level similar to classical UML, providing an overall system view, and low-level modeling, which addresses UML’s limitations in representing quantum algorithms by offering detailed modeling at a lower level of abstraction. This ensures that the quantum system remains both visual and comprehensible in QuanUML. 

The second stage is code generation, where the model is translated into executable code for various quantum programming platforms. The generated code includes quantum operations and control flow derived from the low-level model, as well as classical code corresponding to the high-level model. The details of the code generation process are further discussed in Section \ref{sec:conclusion}.


\subsection{Modeling Quantum Software Systems}
In contrast to existing quantum system modeling techniques, QuanUML provides a comprehensive approach to hybrid quantum-classical system modeling, covering both high-level and low-level perspectives. The high level is derived from classical UML and offers a high-level abstraction of the entire quantum system. Since quantum algorithms are generally concrete with minimal abstraction, QuanUML employs a low-level approach to represent their implementation accurately.

\subsubsection{High Level Modeling}
The high-level approach preserves classical UML while extending it to incorporate quantum systems. From this perspective, hybrid quantum-classical systems are represented as modules that include classical components and quantum algorithms, modeled using QuanUML. 

In the high-level abstraction, the quantum system is modeled using a class diagram, where the entire system is represented as a parent class, and its components are defined as child classes based on design requirements. Classical functions are organized into distinct child classes of the parent system, following classical software engineering principles. For the quantum part, a complete quantum algorithm is modeled as a child class of the entire system, with its lower-level components represented as its child classes. The quantum algorithm class and its child classes are explicitly labeled to distinguish quantum components.

Practically, this labeling is implemented using stereotypes, with \textsf{Quantum} explicitly defined, while no additional labeling is required for classical parts. A specific example is provided in Section \ref{sec:casestudies}.

\subsubsection{Low Level Modeling}
The low-level modeling method focuses on quantum components as independent entities. Unlike classical modeling techniques, which are typically high-level and not algorithm-specific, low-level modeling in QuanUML addresses the detailed nature of quantum algorithms. 
Compared to circuit diagrams, which provide an irreversible representation of the code, a quantum modeling approach based on Model-Driven Development (MDD) is needed to enable transformation from specification to executable code. We propose a low-level modeling approach that aligns with UML sequence diagrams, which are extended in QuanUML to represent quantum state evolution, conditional branching, and classical control. This framework establishes the foundation for future MDD-based code generation.

\begin{figure}
    \centering
    \includegraphics[width=\linewidth]{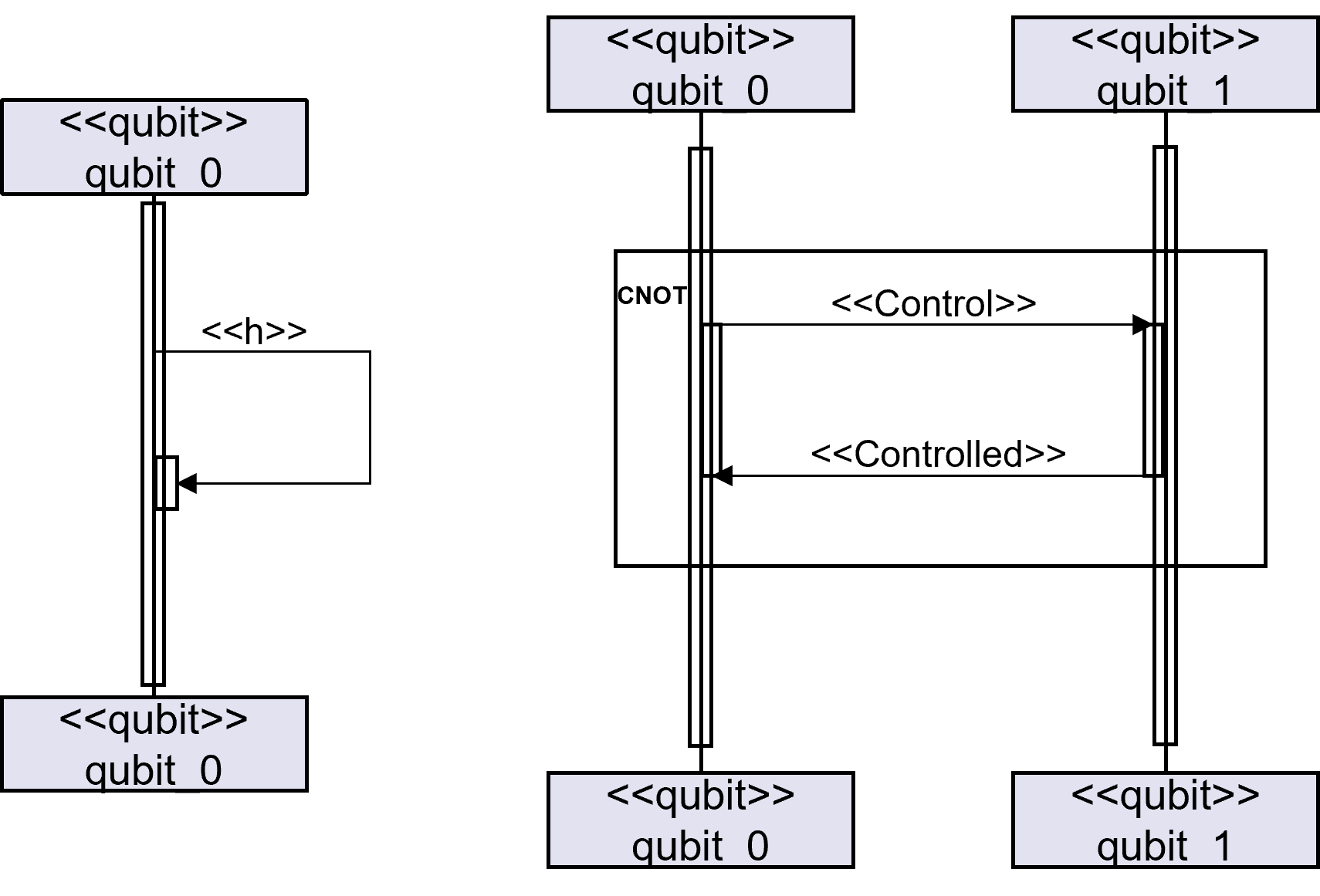}
    \caption{The sequence diagram of quantum gates in QuanUML}
    \label{fig:gate}
\end{figure}

\begin{figure}
    \centering
    \includegraphics[width=0.8\linewidth]{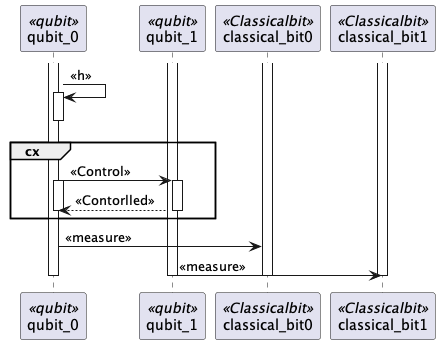}
    \caption{The sequence diagram of Bell State in QuanUML}
    \label{fig:bell_state}
\end{figure}

\begin{itemize}[leftmargin=2em]
    \item \textit{Representation of Qubits and Quantum Gates}: In QuanUML, qubits are represented as distinct entities that, unlike conventional variables, exist in a superposition of multiple states before measurement. 
Quantum gates, which manipulate the state of qubits, are modeled in QuanUML as operations acting on one or more qubits, modifying their states according to the gate's function (e.g., Hadamard, CNOT). Figure \ref{fig:gate} illustrates examples of single-qubit and multi-qubit gates in QuanUML. Qubits are explicitly labeled with the stereotype $\langle\langle \textsf{qubit} \rangle\rangle$ to distinguish them from other entities in a hybrid quantum system. 
The lifeline of a qubit represents its life cycle, from creation to measurement or algorithm completion. It also indicates the life cycle of quantum gates. Single-qubit gates are defined as asynchronous messages due to the absence of control relationships, whereas multi-qubit gates are represented as synchronous and grouping messages to reflect their control relationships and the phenomenon of phase kickback. This representation covers most quantum gates; however, certain specialized gates (e.g., Swap gate) require alternative representations. Section \ref{sec:casestudies} provides illustrative examples.

   \item \textit{Modeling Quantum Superposition and Entanglement}: Quantum superposition and entanglement are among the most challenging aspects of quantum computing modeling, as they involve quantum phenomena absent in classical computing.
In QuanUML, superposition is represented through single-qubit and multi-qubit gates, constrained by the static nature of modeling. Multi-qubit gates are also used to capture all potential entanglement relationships. To precisely depict control relationships and phase kickback, QuanUML employs synchronous messages to represent the control flow between qubits during the gate life cycle. The control flow from the control qubit to the controlled qubit is labeled with the stereotype $\langle\langle \textsf{control} \rangle\rangle$, while phase kickback is labeled with $\langle\langle \textsf{controlled} \rangle\rangle$ or omitted when unnecessary. 
This approach visualizes control and controlled relationships internally, allowing developers to clearly observe how changes in one qubit affect its entangled counterpart. Such visualization is essential for accurately modeling quantum algorithms that rely on entanglement, such as Dynamic Circuits and Shor's Algorithm.

\item \textit{Handling Quantum Measurement and State Transitions}: One of the fundamental differences between classical and quantum computing lies in the measurement process. In classical systems, measurement outcomes are deterministic and depend on the initial state. In contrast, quantum measurements collapse a qubit’s superposition into a deterministic state, with the outcome being probabilistic. 
QuanUML introduces new constructs to model quantum measurements as non-deterministic state transitions. These transitions are depicted in sequence diagrams using lifelines, which conclude the qubit's life cycle at the time of measurement and convey a message indicating the conversion of the measurement result from a quantum state to a classical state. Figure \ref{fig:bell_state} illustrates an example of a Bell state, demonstrating how QuanUML models quantum programs.
QuanUML’s ability to represent these state transitions facilitates the analysis of quantum algorithms and ensures the correctness of quantum state evolution.

\end{itemize}

\subsection{Automated Code Generation}

\subsubsection{Model Transformation from QuanUML to Code}
One of the key benefits of QuanUML is its support for model-driven development, where high-level models are transformed into executable quantum code. QuanUML provides the structure to automate the transformation of quantum models into code compatible with quantum programming frameworks like Qiskit or Cirq. By defining quantum gates, circuits, and qubits at the model level, developers can generate the corresponding quantum instructions without manual coding, reducing errors and increasing productivity. This automated code generation also facilitates rapid prototyping of quantum algorithms, enabling faster iteration cycles in quantum software development.

\subsubsection{Integration with Existing Quantum Frameworks (e.g., Qiskit, Cirq)}
QuanUML is designed to integrate seamlessly with existing quantum programming frameworks, such as Qiskit and Cirq. These frameworks provide the necessary backend to execute quantum algorithms on real quantum hardware or simulators. By generating code that is compatible with these platforms, QuanUML bridges the gap between high-level quantum modeling and low-level quantum programming. This integration allows developers to focus on the logical structure of their quantum programs while QuanUML handles the translation of models into executable code, thus enabling a smooth transition from modeling to implementation.

\section{Case Studies}\label{sec:casestudies}

In this section, we present two case studies to demonstrate the application of QuanUML in modeling quantum algorithms. We have selected two well-known quantum algorithms, Shor's Algorithm and Efficient Long-Range Entanglement using Dynamic Circuit (\textit{Dynamic Circuit} for short), to illustrate how QuanUML models quantum systems and addresses key quantum behaviors, including superposition, entanglement, and measurement. These case studies validate QuanUML’s effectiveness in providing a structured and visual approach to quantum software modeling, facilitating the design and analysis of complex quantum systems.

\subsection{Efficient Long-Range Entanglement using Dynamic Circuit}



\subsubsection{Description of Efficient Long-Range Entanglement using Dynamic Circuit}
Dynamic circuits represent a significant advancement in quantum computing, providing a promising approach for achieving quantum advantage in the near term. They enable the efficient implementation of specific algorithms, such as state teleportation and gate preparation, in the near term and support advanced capabilities like quantum error correction in the long term. Dynamic circuits incorporate classical processing within the qubit’s coherence time, allowing for mid-circuit measurements and feed-forward operations. These operations utilize measurement outcomes to determine the optimal sequence of gates, thus addressing some of the inherent limitations of real hardware.

The following section provides an example that illustrates a six-qubit long-range CNOT gate teleportation. In long-range CNOT gate teleportation, implementing a CNOT gate between the first and last qubits traditionally requires a large number of qubits and numerous swap gates to facilitate interaction, which risks decoherence. However, with a dynamic circuit, only two gates of depth two and a mid-circuit measurement with some classical processing are needed to teleport a CNOT gate.


This dynamic circuit can be effectively modeled in QuanUML by representing each qubit, gate, and operation as elements in sequence diagrams.



\subsubsection{Modeling Dynamic Circuit Using QuanUML}
QuanUML models a dynamic circuit by representing the qubits, quantum gates, and classical processes integrated into a quantum circuit. Existing quantum modeling methods do not represent classical processes within a quantum circuit. To address this, we extended the UML sequence diagram, combining the high-level characteristics of modeling language with the low-level details of a quantum algorithm to represent a dynamic circuit effectively.

\begin{figure*}[t]
  \centering
  \includegraphics[width=0.9\linewidth]{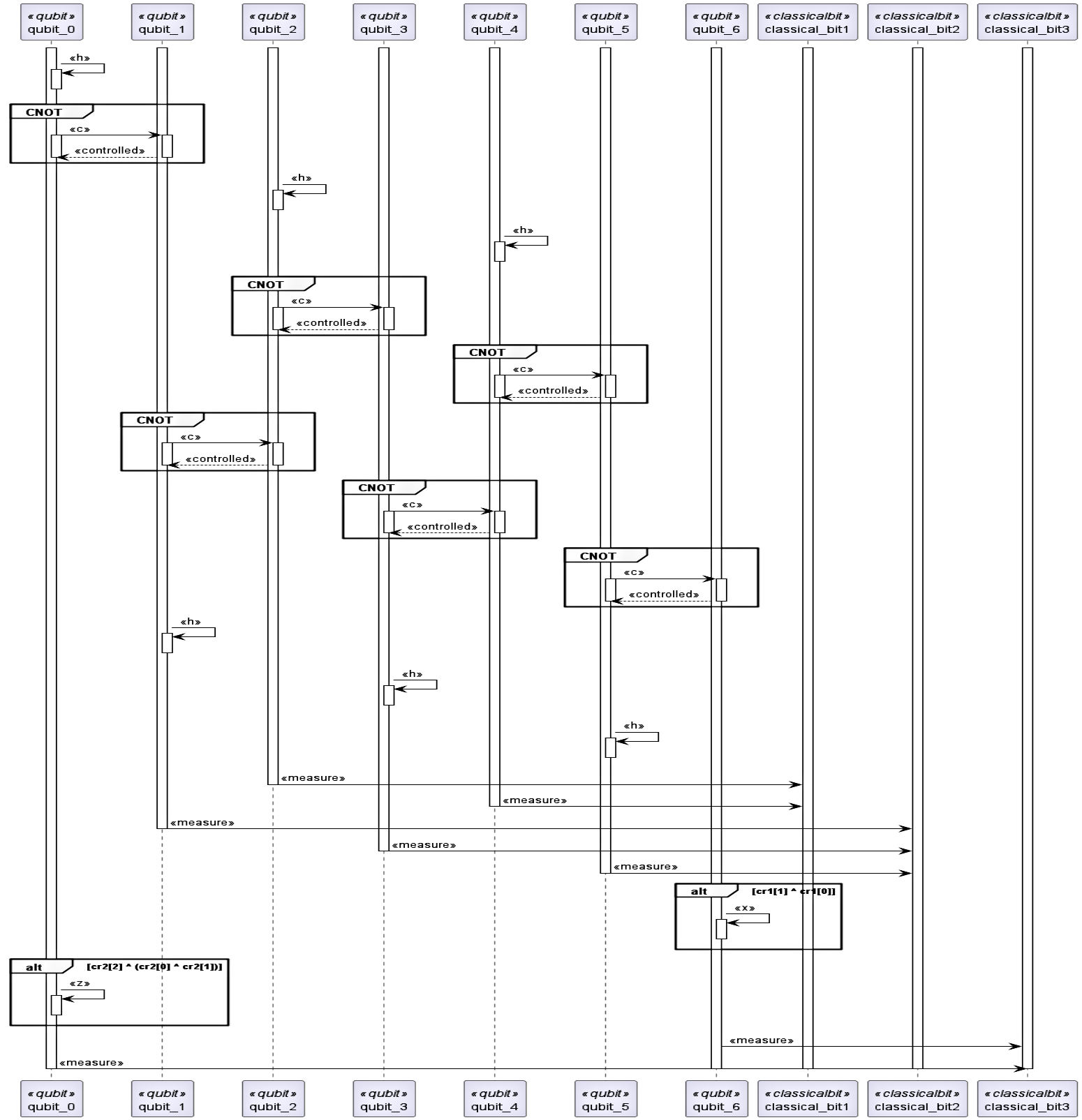}
  \caption[The sequence diagram of Long-Range CNOT Teleportation using dynamic circuit]{The sequence diagram of Long-range CNOT teleportation using dynamic circuit.}
  \label{fig:dynamic}
\end{figure*}

Figure \ref{fig:dynamic} provides an example of a sequence diagram representing Efficient Long-Range Entanglement using a Dynamic Circuit. The left side of the diagram represents the quantum component, including qubit initialization and quantum gate operations, while the right side depicts the classical components. \textsf{Object} represents both quantum and classical bits, with stereotypes "qubit" and "classicalbit" to distinguish them. \textsf{Message} represents gate operations in the quantum circuit; a self-message denotes a single qubit gate named after the quantum gate. Group messages represent multi-quit gates, where the control bit is the sender of the send message, and the controlled bit is the receiver of the send message. The \textsf{Alt} term denotes conditional operations in the classical process, including both conditions and execution steps, while asynchronous messages represent measurement operations.

Additionally, the activation and deactivation of lifelines indicate the creation and collapse of qubits, classical bits, and quantum gates. Thus, users can trace the life cycle of each component within the quantum circuit. This is a characteristic inherent to sequence diagrams.

\subsubsection{Advantages of Modeling Dynamic Circuit with QuanUML}
QuanUML provides a clear and structured approach to modeling dynamic circuits. It lets developers visualize the modeling process and understand how classical processes influence quantum components. Developers gain a more intuitive understanding of circuit behavior using sequence diagrams for gate operations. Additionally, QuanUML accurately represents the collapsing nature of quantum measurement, allowing developers to assess qubit states. This visual modeling simplifies the complex task of designing and analyzing dynamic circuits, making it accessible to both quantum and classical software developers.

\subsection{Shor's Algorithm}
\subsubsection{Description of Shor's Algorithm}

Shor's Algorithm factors a large integer by combining quantum and classical processes. The algorithm is most notably known for its ability to efficiently factor large numbers, posing a potential threat to classical encryption methods such as RSA. The quantum portion of Shor’s Algorithm involves creating a superposition of quantum states, applying modular exponentiation, and then performing a quantum Fourier transform to extract periodicity information, which can be used in the classical post-processing phase to factor the integer \cite{shor1994algorithms}. The following pseudocode (Algorithm 1 outlines the quantum portion of Shor's Algorithm:

\begin{algorithm}
\caption{Shor's Algorithm}[Shor’s Algorithm]
\begin{algorithmic}[1]
    \State If a composite number $N$ is even, return factor 2.
    \State Determine whether $N = a^b$ for integers $a \geq 1$ and $b \geq 2$, and if so return the factor $a$.
    \State Randomly choose $x$ in the range $1$ to $N - 1$. If $gcd(x, N) > 1$ then return the factor $gcd(X, N)$.
    \State Use the order-finding subroutine to find the order $r$ of $x$ modulo $N$.
    \State If $r$ is even and $x^{r/2} \neq -1 (mod N)$ then compute $gcd(x^{r/2} - 1, N)$ and $gcd(x^{r/2} + 1, N)$, and test to see if one of these is a non-trivial factor, returning that factor if so. Otherwise, the algorithm fails.
\end{algorithmic}
\end{algorithm}


The primary challenge in Shor's Algorithm is determining the period of $f(x) = a^x \mod N$. Shor's Algorithm addresses this using a quantum subroutine, often called the period-finding subroutine, which requires constructing a specific circuit. In this example, we select $N = 15$ without limiting the choice of $a$. While one particular $a$ with prior information could allow for a more concise, specialized period-finding circuit, relying heavily on such information limits versatility. Therefore, we avoid this simplified approach. QuanUML enables the specification to extend to all permissible values of $a$ for the case of $N = 15$.



\subsubsection{Modeling Shor's Algorithm using QuanUML}
The large number of quantum gates involved makes the modeling process cumbersome. In quantum programming, a common approach to mitigate this issue is using custom gates, which enable the reuse of groups of gates with similar functions. However, previous quantum modeling methods have primarily overlooked this approach, further complicating the modeling process.

QuanUML allows the modeling of custom sub-quantum algorithms to simplify the modeling process at a low level. A sequence diagram represents the quantum component, dividing the circuit into sub-circuits such as the overall circuit, period finding, and QFT dagger. At the high level, a class diagram represents the components and relationships within Shor's algorithm, viewing it as a complete system rather than a simple quantum circuit. The QuanUML modeling method described here aligns with the high-level characteristics of the model-driven approach while incorporating the low-level details of the quantum algorithm.

\begin{figure}[h]
  \centering
\includegraphics[width=\linewidth]{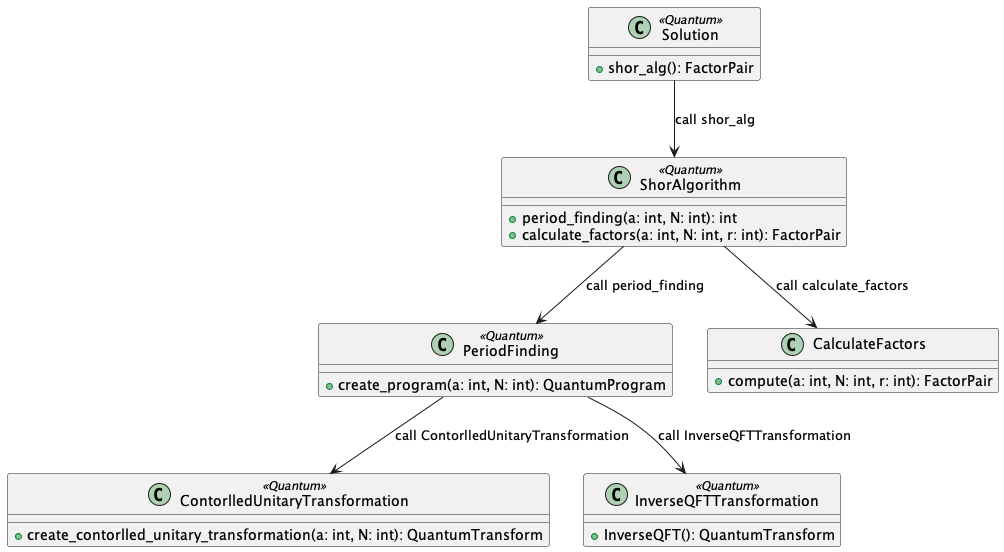}
  \caption[Class Diagram of Shor Algorithm]{The class diagram of Shor's Algorithm in QuanUML.}
  \label{fig:shor class}
\end{figure}

Figure \ref{fig:shor class} illustrates a class diagram outlining the architectural framework of the Shor algorithm for modulus $N = 15$. The stereotype "$\langle\langle \textsf{Quantum} \rangle\rangle$" denotes all quantum circuits and classes containing quantum circuits, distinguishing quantum components from classical ones in the class diagram. The \textsf{Solution} class is responsible for invoking the Shor algorithm, while the \textsf{ShorAlgorithm} class, serving as the main class, handles the factorization process. This involves calling the \textsf{PeriodFinding} class to compute the period and the \textsf{CalculateFactor} class to determine the factors. 

\begin{figure*}[htb]
  \centering
  \includegraphics[width=0.95\linewidth]{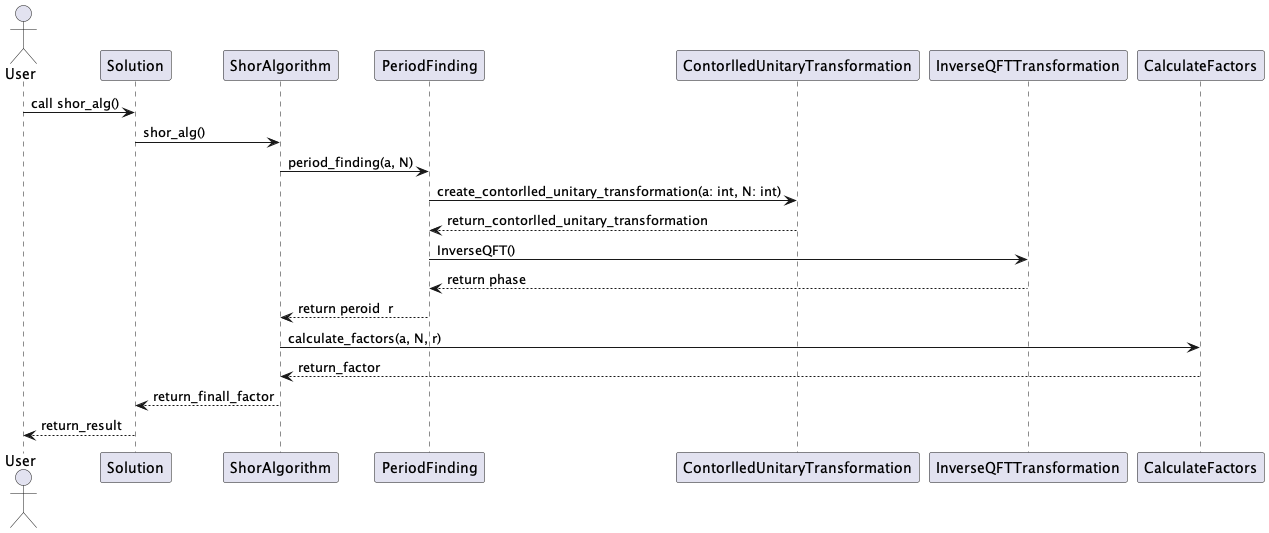}
  \caption[Sequence Diagram of Shor Algorithm]{Abstract sequence diagram of Shor's Algorithm.}
  \label{fig:shor Sequence Diagram}
\end{figure*}

The \textsf{ControlledUnitaryTransformation} function generates the quantum unitary transformation of the Shor algorithm based on the input $a$ and $N$. Similarly, the \textsf{InverseQFTFunction} creates the inverse QFT circuit corresponding to $a$ and returns it to \textsf{PeriodFinding}. Furthermore, Figure \ref{fig:shor Sequence Diagram} illustrates the system's operational methodology.

\begin{figure*}[th]
  \centering
  \includegraphics[width=0.95\linewidth]{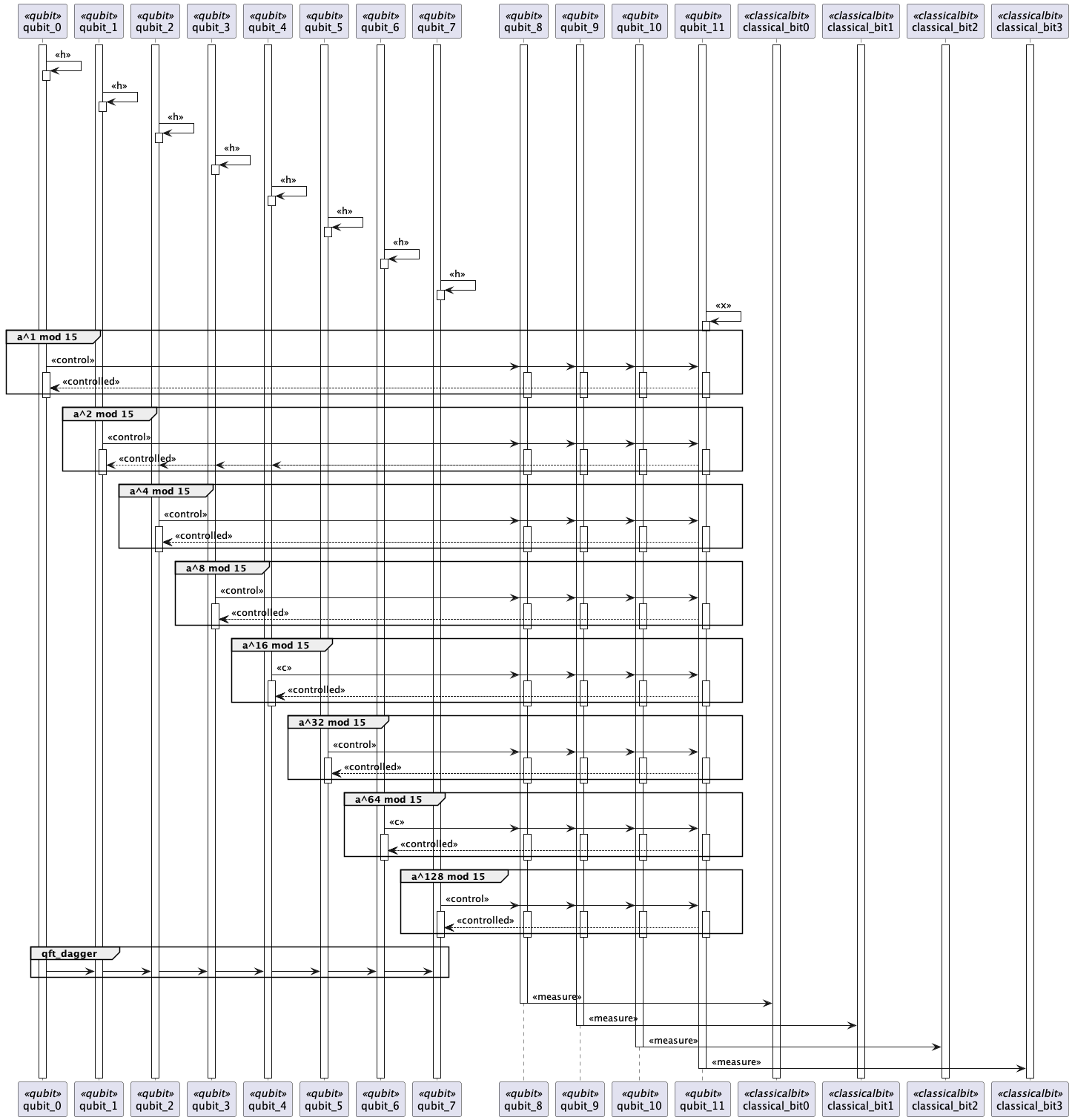}  \caption[Quantum Circuit of Shor Algorithm]{Low level sequence diagram of Shor's Algorithm.}
  \label{fig:shor integrated}
\end{figure*}

\begin{figure}[h]
  \centering
  \includegraphics[width=0.7\linewidth]{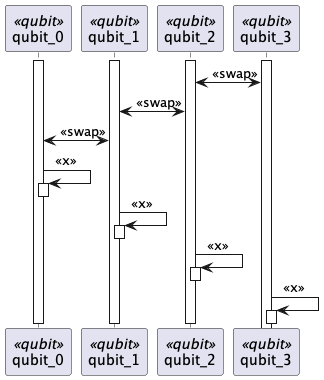}
  \caption[Period Finding for 7 mod 15]{The sequence diagram of period finding for 7 mod 15.}
  \label{fig:shor 7mod15}
\end{figure}

Figure \ref{fig:shor integrated} illustrates the configuration of the Shor algorithm mod 15, integrating high-level and low-level elements to enhance the conciseness of the QuanUML design. The QFT dagger and period-finding components represent high-level elements, clarifying the overall design structure of the Shor algorithm’s quantum component. Figure \ref{fig:shor 7mod15} illustrates the detailed low-level design for period finding specific to $7 \mod 15$. This section presents the low-level implementation in abstract form within the overall circuit, making the QuanUML representation more streamlined and concise. Combining high-level and low-level elements, this approach streamlines the design process and aligns QuanUML with the model-driven design concept.

QuanUML’s visual representations of these processes make it easier to understand the interplay between quantum and classical components in Shor’s Algorithm.

\subsubsection{Challenges and Insights from the Case Study}
Modeling Shor’s Algorithm using QuanUML highlights the complexity of hybrid quantum-classical algorithms. QuanUML’s ability to integrate classical and quantum components in a single model proves to be beneficial for visualizing the entire algorithm workflow. However, the complexity of the quantum Fourier transform and the modular exponentiation steps presents challenges in representing every quantum state explicitly due to the exponential growth in the number of states \cite{volovich2001quantum}. Despite these challenges, QuanUML provides an effective framework for visualizing the key components of Shor’s Algorithm, facilitating both analysis and potential optimizations.

\subsection{Comparison}
We compare QuanUML with prior works in terms of functionality and efficiency. For the functional comparison, we evaluate QuanUML against classical UML and UML profile-based modeling methods~\cite{perez2022design} across static circuits, dynamic circuits, and Shor's algorithm, considering both high- and low-level representations. The results in Table~\ref{tab:functionCompare} show that classical UML fails to support one of the tasks, while profile-based UML is limited to static circuits and Shor's algorithm. In contrast, QuanUML successfully handles all tasks, demonstrating its broader applicability.

For efficiency, we compare the number of elements in QuanUML with those in the UML profile-based modeling method using several basic quantum algorithms previously implemented with the UML profile-based approach. The results in Table~\ref{tab:efficiency} indicate that QuanUML requires fewer elements while maintaining the same performance. Specifically, in the 2-qubit Grover algorithm, QuanUML reduces the number of elements by nearly half compared to the UML profile-based modeling method. However, in the 4-qubit Full Adder algorithm, both methods use a similar number of elements.

To explain this discrepancy, we conducted an in-depth analysis and found that QuanUML requires fewer elements for single-qubit gates. Additionally, in QuanUML, the number of elements for multi-qubit gates increases with the complexity of the control relationships within the gate. In contrast, in the UML profile-based modeling method, the element count depends only on the number of qubits in the gate. This difference arises because QuanUML more precisely represents control relationships in multi-qubit gates. Both the GHZ and Full Adder algorithms utilize CCNOT gates, but the Full Adder algorithm requires two CCNOT gates, leading to a higher number of elements in QuanUML.

\begin{table}
    \caption{Function comparison with previous work~\cite{perez2022design}}
    \label{tab:functionCompare}    \centering
    \includegraphics[width=\linewidth]{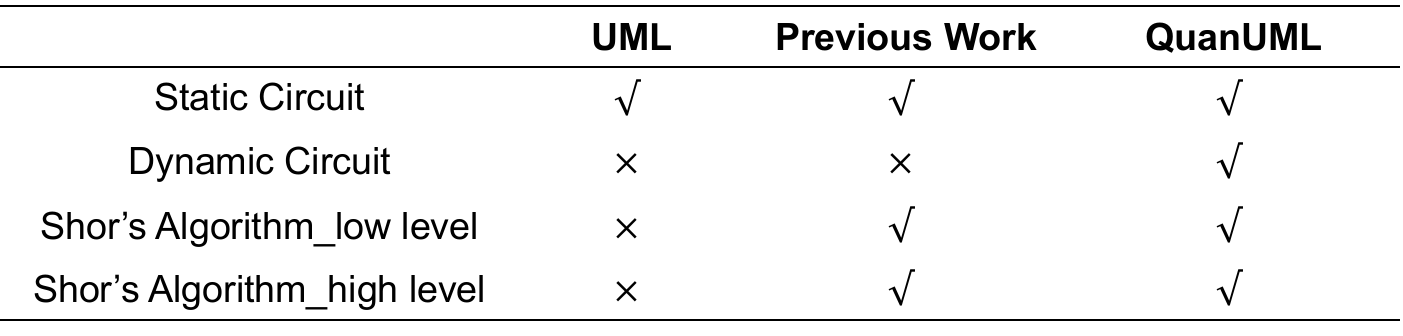}
\end{table}

\begin{table}[]
    \caption{Element comparison with previous work~\cite{perez2022design}}
    \label{tab:efficiency}
    \centering
    \includegraphics[width=\linewidth]{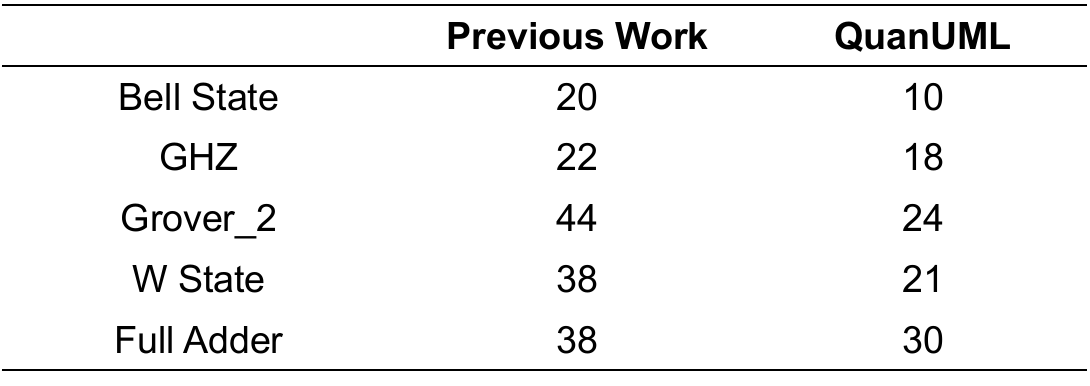}
\end{table}

\subsection{Discussions}
The Efficient Long-range entanglement case studies using a Dynamic Circuit and Shor's Algorithm demonstrate QuanUML's ability to effectively model various quantum behaviors, including superposition, entanglement, measurement, and quantum-classical interaction.
 QuanUML offers a structured and visual approach to quantum software modeling, making it easier for developers to design and analyze quantum systems. However, as quantum algorithms become more complex, the scalability of QuanUML will be tested, particularly in representing large quantum circuits and highly entangled systems. Future work should explore ways to optimize the visualization of such complex systems and integrate more advanced modeling techniques to handle the growing complexity of quantum algorithms.

\section{Related Work}\label{sec:relatedwork}
This section reviews key approaches for modeling quantum software.

Pérez-Delgado and Perez-Gonzalez~\cite{Perez-Delgado2020quantum} propose Q-UML, an extension of UML for modeling quantum algorithms. Their approach integrates quantum-specific constructs, such as qubits and quantum gates, into UML diagrams while maintaining compatibility with classical UML. Q-UML organizes quantum programs using class diagrams to define discrete modules and sequence diagrams to illustrate interactions between them. Their primary focus is on hybrid quantum-classical systems, ensuring a unified modeling framework for both components.
QuanUML is comparable to Q-UML in high-level representation but extends capabilities to low-level modeling, addressing the limitations of classical UML in representing quantum algorithms. It provides a comprehensive framework tailored for model-driven quantum software development, supporting not only quantum circuits but also broader quantum software architectures. QuanUML introduces a richer set of abstractions, enabling detailed modeling beyond the hybridization approach of Q-UML.

Pérez-Castillo et al.~\cite{perez2022design} develop the Quantum UML Profile, extending UML to incorporate quantum elements like qubits, gates, and measurements. Their approach integrates quantum constructs into UML activity and class diagrams, facilitating the representation of complex quantum circuits, such as quantum teleportation. By leveraging UML’s profile mechanism, their framework enables visual modeling of quantum algorithms and automatic translation into executable Qiskit code.
Compared to QuanUML, the approach by Pérez-Castillo et al. follows a more classical paradigm, embedding quantum algorithms as modules within UML rather than establishing a true classical-quantum hybrid. While both approaches extend UML for quantum modeling, QuanUML aims for a broader software modeling framework, covering a wider range of quantum systems. In contrast, the Quantum UML Profile specializes in quantum circuit representation and Qiskit code generation, making it more suited for algorithm-specific modeling.

Ali and Yue~\cite{ali2020modeling} introduce a platform-independent framework that abstracts core quantum concepts, such as quantum states, entanglement, and operations, into a conceptual model. Their framework provides high-level abstractions to bridge the gap between classical software engineering and quantum programming, remaining independent of specific quantum languages or hardware, making it broadly applicable.
While Ali and Yue’s approach supports high-level modeling, it lacks concrete implementation details, particularly when interfacing with specific quantum hardware or languages. The framework remains abstract, limiting its direct applicability to practical quantum software development without further refinement or integration with specific tools.

\section{Conclusion and Future Work}\label{sec:conclusion}

In this paper, we introduced QuanUML, an extension of UML designed specifically for modeling quantum software systems. QuanUML incorporates quantum-specific constructs such as qubits, quantum gates, and quantum circuits, making it suitable for modeling quantum algorithms. We demonstrated how QuanUML can be applied to Dynamic Circuit and Shor's Algorithms, showing its ability to model key quantum behaviors like superposition, entanglement, and measurement.

QuanUML supports model-driven development for quantum software, enabling high-level modeling and automated code generation. This approach reduces the complexity of quantum software design and facilitates the development and analysis of quantum programs. While QuanUML provides a structured framework for modeling quantum algorithms, further work is needed to improve scalability and support more complex quantum systems. Code generation remains a key area for future work, and we plan to extend QuanUML’s capabilities to support multiple quantum programming languages, including Qiskit~\cite{qiskit2024}, Q\#~\cite{svore2018q}, Cirq~\cite{cirq2018google}, and Braket~\cite{braket}. Additionally, future research should focus on expanding QuanUML to accommodate a broader range of quantum algorithms and integrating it with additional quantum programming frameworks to enhance its applicability in real-world quantum software development.

\bibliographystyle{IEEEtran}
\bibliography{qse-bibliography}
\end{document}